\documentclass[letterpaper, twocolumn, superscriptaddress, floatfix]{revtex4}
\pdfoutput=1
\usepackage{xspace}
\usepackage{amsmath}
\usepackage{graphicx}

\usepackage[usenames]{color}

\newcommand{\ChiSq}{\ensuremath{\chi^2}\xspace}
\newcommand{\HchiSq}{\ensuremath{H^{\chi^2}}\xspace}

\newcommand{\BrownEtAl}{Brown \emph{et~al}.\xspace}

\newcommand{\citecoll}{\cite{bib:Tyson1991, bib:Zwolak2005a, bib:Goldbeter1991, bib:Vilar2002, bib:Edelstein1996, bib:Kholodenko2000, bib:Lee2003, bib:Leloup1999, bib:Brown2004, bib:Dassow2000, bib:Ueda2001, bib:Locke2005, bib:Zak2003, bib:Curto1998, bib:Chassagnole2002, bib:Chen2004, bib:Sasagawa2005}\xspace}

\begin{document}
%\vskip\abovecaptionskip
%\hbox to \hsize{\hfil #1\hfil}%
%\vskip\belowcaptionskip}

\date{\today}

\pagestyle{myheadings}
\markboth{}{Sloppy Systems Biology}

% Needs to be 75 characters or less
\title{Universally Sloppy Parameter Sensitivities in Systems Biology Models}

\author{Ryan N. Gutenkunst}
\email{rng7@cornell.edu}
\affiliation{Laboratory of Atomic and Solid State Physics, Cornell University, Ithaca, NY, USA}
\author{Joshua J. Waterfall}
\affiliation{Department of Molecular Biology and Genetics, Cornell University, Ithaca, NY, USA}
\author{Fergal P. Casey}
\affiliation{UCD Conway Institute of Biomolecular \& Biomedical Research, University College Dublin, Belfield, Ireland}
\author{Kevin S. Brown}
\affiliation{Department of Molecular and Cellular Biology, Harvard University, Cambridge, MA, USA}
\author{Christopher R. Myers}
\affiliation{Cornell Theory Center, Cornell University, Ithaca, NY, USA}
\author{James P. Sethna}
\affiliation{Laboratory of Atomic and Solid State Physics, Cornell University, Ithaca, NY, USA}

% 250 - 300 words
\begin{abstract}
Quantitative computational models play an increasingly important role in modern biology.  
Such models typically involve many free parameters, and assigning their values is often a substantial obstacle to model development.
Directly measuring \emph{in vivo} biochemical parameters is difficult, and collectively fitting them to other experimental data often yields large parameter uncertainties.
Nevertheless, in earlier work we showed in a growth-factor-signaling model that collective fitting could yield well-constrained predictions, even when it left individual parameters very poorly constrained.
We also showed that the model had a `sloppy' spectrum of parameter sensitivities, with eigenvalues roughly evenly distributed over many decades.
Here we use a collection of models from the literature to test whether such sloppy spectra are common in systems biology.
Strikingly, we find that every model we examine has a sloppy spectrum of sensitivities.
We also test several consequences of this sloppiness for building predictive models.
In particular, sloppiness suggests that collective fits to even large amounts of ideal time-series data will often leave many parameters poorly constrained. Tests over our model collection are consistent with this suggestion.
This difficulty with collective fits may seem to argue for direct parameter measurements, but sloppiness also implies that such measurements must be formidably precise and complete to usefully constrain many model predictions.
We confirm this implication in our growth-factor-signaling model.
Our results suggest that sloppy sensitivity spectra are universal in systems biology models.
The prevalence of sloppiness highlights the power of collective fits and suggests that modelers should focus on predictions rather than on parameters.
\end{abstract}

\maketitle

\textbf{Non-technical Summary:}
Dynamic systems biology models typically involve many kinetic parameters, the quantitative determination of which has been a serious obstacle to using these models.
Previous work showed for a particular model that useful predictions could be extracted from a fit long before the experimental data constrained the parameters, even to within orders of magnitude.
This was attributed to a `sloppy' pattern in the model's parameter sensitivities; the sensitivity eigenvalues were roughly evenly spaced over many decades. 
Consequently, the model behavior depended effectively on only a few `stiff' parameter combinations.
Here we study the converse problem, showing that direct parameter measurements are very inefficient at constraining the model's behavior.
To yield effective predictions such measurements must be very precise and complete; even a single imprecise parameter often destroys predictivity. 
We also show here that the characteristic sloppy eigenvalue pattern is reproduced in sixteen other diverse models from the systems biology literature. 
The apparent universality of sloppiness suggests that predictions from most models will be very fragile to single uncertain parameters and that collective parameters fits can often yield tight predictions with loose parameters.
Together these results argue that focusing on parameter values may be a very inefficient route to useful models.

\noindent\hrulefill

Dynamic computational models are powerful tools for developing and testing
hypotheses about complex biological systems~\cite{bib:Kitano2002,
bib:Locke2005, bib:Voit2006}.  It has even been suggested that such models will
soon replace databases as the primary means for exchanging biological
knowledge~\cite{bib:Aldridge2006}.  A major challenge with such models,
however, is that they often possess tens or even hundreds of free parameters
whose values can significantly affect model
behavior~\cite{bib:Ingram2006,bib:Mayo2006}.  While high-throughput methods for
discovering interactions are well-developed~\cite{bib:Sachs2005},
high-throughput methods for measuring biochemical parameters remain
limited~\cite{bib:Maerkl2007}.  Furthermore, using values measured \emph{in vitro} in an \emph{in vivo} application may introduce substantial
inaccuracies~\cite{bib:Minton2001, bib:Teusink2000}.  On the other hand,
collectively fitting parameters~\cite{bib:Mendes1998,bib:Jaqaman2006} by
optimizing the agreement between the model and available data often yields
large parameter uncertainties~\cite{bib:Brodersen1987,bib:Cho2003,bib:Rodriguez-Fernandez2006}.
In approaches typically more focused on steady-state distributions of 
fluxes in metabolic networks, metabolic control analysis has been used 
to quantify the sensitivity of model behavior with respect to
parameter variation~\cite{bib:Fell1997}, and flux-balance analysis and related techniques have probed the robustness of metabolic networks~\cite{bib:Wiback2004, bib:Famili2005}.

One way to cope with the dearth of reliable parameter values is to focus on
predictions that are manifestly parameter-independent~\cite{bib:Bailey2001},
but these are mostly qualitative.  An alternative is not to forsake
quantitative predictions, but to accompany them with well-founded
uncertainty estimates based on an ensemble of parameter sets statistically drawn from all sets
consistent with the available data~\cite{bib:Brown2003a}.
(Uncertainties in the model structure itself may be important in some cases.
Here we focus on parameter uncertainties, as they are often important on their
own.)

Brown \emph{et~al}.\ took this approach in developing a computational model of the well-studied growth-factor-signaling network in PC12
cells~\cite{bib:Brown2004}.  They collectively fit their model's 48 parameters to 68 data points from 14 cell-biology experiments (mostly Western blots).  After the fit, all 48
parameters had large uncertainties; their 95\% confidence intervals each
spanned more than a factor of 50.  Surprisingly, while fitting this modest
amount of data did not tightly constrain any single parameter value, it did
enable usefully tight quantitative predictions of behavior under interventions,
some of which were verified experimentally.

In calculating their uncertainties, \BrownEtAl found that the quantitative
behavior of their model was much more sensitive to changes in certain
combinations of parameters than others.  Moreover, the sensitivity eigenvalues were
approximately equally spaced in their logarithm, a pattern deemed `sloppy'.
Such sloppy sensitivities were subsequently seen in other multi-parameter
fitting problems, from interatomic potentials~\cite{bib:Frederiksen2004} to
sums of exponentials~\cite{bib:Waterfall2006}.  The fact that sloppiness arises in such disparate contexts suggests that it
may be a universal property of nonlinear multi-parameter models. (Here the term `universal' has a technical meaning from statistical physics, denoting a shared common property with a deep underlying cause; see~\cite{bib:Waterfall2006}.
Universality in this sense does not imply that all models must necessarily share the property.)
%For example, a system whose experiments are concocted to constrain each parameter separately would not be sloppy.

In this work, we begin by empirically testing seventeen systems biology
models from the literature, examining the sensitivity of their behavior to
parameter changes. Strikingly, we find that Brown \emph{et~al}'s model is not 
unique in its sloppiness; every model we examine exhibits a sloppy parameter
sensitivity spectrum. (Thus, in the models we've examined sloppiness is also
universal in the common English sense of ubiquity.) We
then study the implications of sloppiness for 
constraining parameters and predictions. We argue that obtaining precise
parameter values from collective fits will remain difficult even with extensive
time-series data, because the behavior of a sloppy model is very insensitive
to many parameter combinations. We also argue that, to usefully constrain model
predictions, direct parameter measurements must be both very precise and
complete, because sloppy models are also conversely very sensitive to some parameter 
combinations. Tests over our collection of models support the first
prediction, and detailed analysis of the model of \BrownEtAl supports the second contention.

Sloppiness, while not unique to biology, is particularly relevant to biology,
because the collective behavior of most biological systems is much
easier to measure \emph{in vivo} than the values of individual parameters. 
Much work has focused on optimizing experimental design to best constrain model parameters with collective fits~\cite{bib:Faller2003,bib:Zak2003,bib:Gadkar2005}.
% or large programs to measure such parameters directly~\cite{XXX}.
We argue against this focus on parameter values, 
particularly when our understanding of a system is tentative and incomplete. Concrete predictions can be extracted from models long before their parameters are even roughly known~\cite{bib:Brown2004}, and
when a system is not already well-understood, it can be more profitable to design experiments to directly improve predictions of interesting system behavior~\cite{bib:Casey2006} rather than to improve estimates of parameters.

\section{Results}

\subsection{Systems Biology Models have Sloppy Sensitivity Spectra}
\label{sec:SloppySpectra}

Our collection of 17 systems biology models~\citecoll was drawn primarily from
the BioModels database~\cite{bib:BioModels}, an online repository of models
encoded in the Systems Biology Markup Language (SBML)~\cite{bib:Hucka2003}.
The collected models encompass a diverse range of biological systems, including
circadian rhythm, metabolism, and signaling.  All the models are formulated as
systems of ordinary differential equations, and they range from having about ten to
more than two hundred parameters.  In most cases, these parameters were not
systematically fit or measured in the original publication.

We quantified the change in model behavior as parameters $\theta$ varied from
their published values $\theta^*$ by the average squared change in molecular
species time courses: 
\begin{equation} \ChiSq(\theta) \equiv
\frac{1}{2\,N_c\,N_s} \sum_{s, c}\frac{1}{T_c} \int^{T_c}_0
\left[\frac{y_{s,c}(\theta, t) - y_{s,c}(\theta^*,
t)}{\sigma_s}\right]^2\,\mathrm{d}t,\label{eqn:ChiSq} 
\end{equation}
a kind of continuous least-squares fit of parameters $\theta$ to `data' 
simulated from published parameters $\theta^*$.
Here $y_{s,c}(\theta, t)$ is the time course of molecular species $s$ given
parameters $\theta$ in condition $c$, and $T_c$ is the `measurement' time for
that condition.  We took the species normalization $\sigma_s$ to be equal to
the maximum value of species $s$ across the conditions considered; other consistent normalizations yield the same qualitative conclusions.

For each model, the
sum in Equation~\ref{eqn:ChiSq} runs over all molecular species in the model
and (except where infeasible) over all experimental conditions considered in
the corresponding paper---an attempt to neutrally measure system behavior under
conditions deemed significant by the original authors. (The total number
of conditions and species are denoted by $N_c$ and $N_s$, respectively.) 
SBML files and SloppyCell~\cite{bib:SloppyCell}
scripts for all models and conditions are available in Dataset S1.

\begin{figure*} 
\begin{center} 
\includegraphics{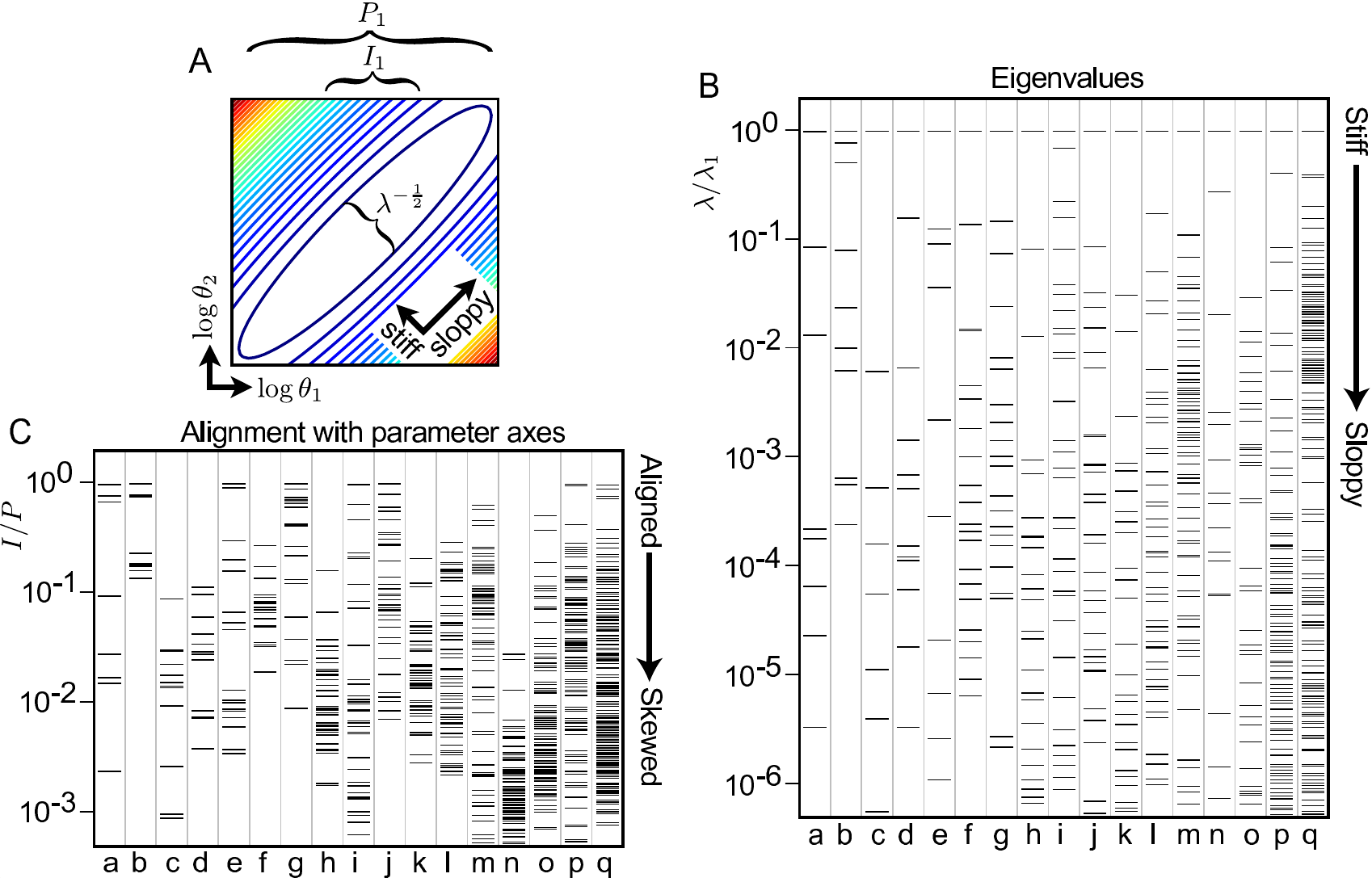} 
\caption{Parameter sensitivity spectra\\
Subfigure A illustrates the quantities we calculate from \HchiSq, while subfigures B and C show that all the models we examined have sloppy sensitivity spectra.\\
A: Analyzing \HchiSq corresponds to approximating the surfaces of constant
model behavior change (constant \ChiSq)
as ellipsoids. The width of each principal axis is proportional to one over the
square root of the corresponding eigenvalue. The inner ellipsoid's projection
onto and intersection with the $\log \theta_1$ axis are denoted $P_1$
and $I_1$, respectively.\\
B: Plotted are the eigenvalue spectra of \HchiSq for our collection of systems biology models. The many decades generally spanned indicate the ellipses have very large aspect ratio. (The spectra have each been normalized by their largest eigenvalue. Not all values are visible for all models.)\\
C: Plotted is the spectrum of $I/P$ for each parameter
in each model in our collection. Generally very few parameters have $I/P \approx 1$, suggesting that the ellipses are skewed from the bare parameter axes. (Not all values are visible for all models.)\\
The models are ordered by increasing number of free parameters and are: 
(a) eukaryotic cell cycle~\protect\cite{bib:Tyson1991}, 
(b) Xenopus egg cell cycle~\protect\cite{bib:Zwolak2005a},
(c) eukaryotic mitosis~\protect\cite{bib:Goldbeter1991},
(d) generic circadian rhythm~\protect\cite{bib:Vilar2002},
(e) nicotinic acetylcholine intra-receptor dynamics~\protect\cite{bib:Edelstein1996},
(f) generic kinase cascade~\protect\cite{bib:Kholodenko2000},
(g) Xenopus Wnt signaling~\protect\cite{bib:Lee2003},
(h) Drosophila circadian rhythm~\protect\cite{bib:Leloup1999},
(i) rat growth-factor signaling~\protect\cite{bib:Brown2004},
(j) Drosophila segment polarity~\protect\cite{bib:Dassow2000},
(k) Drosophila circadian rhythm~\protect\cite{bib:Ueda2001},
(l) Arabidopsis circadian rhythm~\protect\cite{bib:Locke2005},
(m) \emph{in silico} regulatory network~\protect\cite{bib:Zak2003},
(n) human purine metabolism~\protect\cite{bib:Curto1998},
(o) E. coli carbon metabolism~\protect\cite{bib:Chassagnole2002},
(p) budding yeast cell cycle~\protect\cite{bib:Chen2004},
(q) rat growth-factor signaling~\protect\cite{bib:Sasagawa2005}.}
\label{fig:sloppiness} \end{center}
\end{figure*}

To analyze each model's sensitivity to parameter variation, we considered the
Hessian matrix corresponding to \ChiSq: 
\begin{equation} 
\HchiSq_{j,k} \equiv
\frac{d^2 \chi^2(\theta)}{d \log \theta_j\,d \log
\theta_k}.\label{eqn:HchiSq} 
\end{equation} 
We took derivatives with respect to
$\log \theta$ to consider \emph{relative} changes in parameter values, because
biochemical parameters can have different units and widely varying scales.
Analyzing \HchiSq corresponds to approximating the surfaces of constant model
behavior deviation (as quantified by \ChiSq) to be $N_p$-dimensional
ellipsoids, where $N_p$ is the number of parameters in the model.
Figure~\ref{fig:sloppiness}A schematically illustrates these ellipsoids and
some of their characteristics.  (Details of calculating \HchiSq and related
quantities are found in Methods. Dataset S1 includes \HchiSq for each model.)

The principal axes of the ellipsoids are the eigenvectors of \HchiSq, and the
width of the ellipsoids along each principal axis is proportional to one over
the square root of the corresponding eigenvalue.  The narrowest axes are called
`stiff', and the broadest axes `sloppy'~\cite{bib:Brown2003a}.  The eigenvalue
spectra for the models in our collection are shown in
Figure~\ref{fig:sloppiness}B (each normalized by its largest eigenvalue).  In
every case, the eigenvalues span many decades.  All but one span more than $10^6$,
indicating that the sloppiest axes of the ellipsoids illustrated in
Figure~\ref{fig:sloppiness}A are generally more than one thousand times as long as the
stiffest axes.  In each spectrum the eigenvalues are also approximately evenly
spaced in their logarithm; there is no well-defined cutoff between `important'
and `unimportant' parameter combinations.

The Hessian matrix is a local quadratic approximation to the generally nonlinear \ChiSq function. Principal component analysis of extensive Monte Carlo runs in the \BrownEtAl model, however, indicates that the sloppiness revealed by \HchiSq is indicative of full nonlinear \ChiSq function~\cite{bib:Brown2003a}.

Along with their relative widths, the degree to which the principal axes of the
ellipsoids are aligned to the bare parameter axes is also important.  We
estimated this by comparing the ellipsoids' intersections $I_i$ with
and projections $P_i$ onto each bare parameter axis $i$.  If
$I_i/P_i = 1$ then one of the principal axes of the
ellipsoids lies along bare parameter direction $i$.
Figure~\ref{fig:sloppiness}C plots the $I/P$ spectrum for each
model.  In general, very few axes have $I/P \approx 1$; the
ellipses are skewed from single parameter directions.

Naively, one might expect the stiff eigenvectors to embody the most
important parameters and the sloppy directions to embody parameter correlations
that might suggest removable degrees of freedom, simplifying the model.
Empirically, we have found that the eigenvectors often tend to involve significant components of many different parameters; plots of the five stiffest eigenvectors for each model are in Supporting Text S1.
This is understandable theoretically; the nearly-degenerate sloppy eigenvectors  should mix, and the stiff eigenvectors can include arbitrary admixtures of unimportant directions to a given important parameter combination.
(Indeed, in analogous random-matrix theories the eigenvectors are known to be uncorrelated random vectors~\cite{bib:Mehta2004}.) 
While the relatively random eigenvectors studied here may not be useful
in guiding model reduction, more direct explorations of parameter 
correlations have yielded interesting correlated parameter
clusters~\cite{bib:WaterfallPhD}.
%Here we focus on model behavior; other approaches to model reduction are being pursued elsewhere~\cite{bib:WaterfallPhD}.

These characteristic parameter sensitivities that evenly span many decades and are skewed from bare parameter axes
define a `sloppy' model~\cite{bib:Brown2003a}.
Figures~\ref{fig:sloppiness}B and~\ref{fig:sloppiness}C show that every model we have examined has a sloppy sensitivity spectrum.  Next we discuss some broad questions about the relation between model predictions, collective fits, and parameter measurements and see how the sloppy properties of these models may suggest answers.

\subsection{Consequences of Sloppiness} 

\begin{figure} \begin{center}
\includegraphics{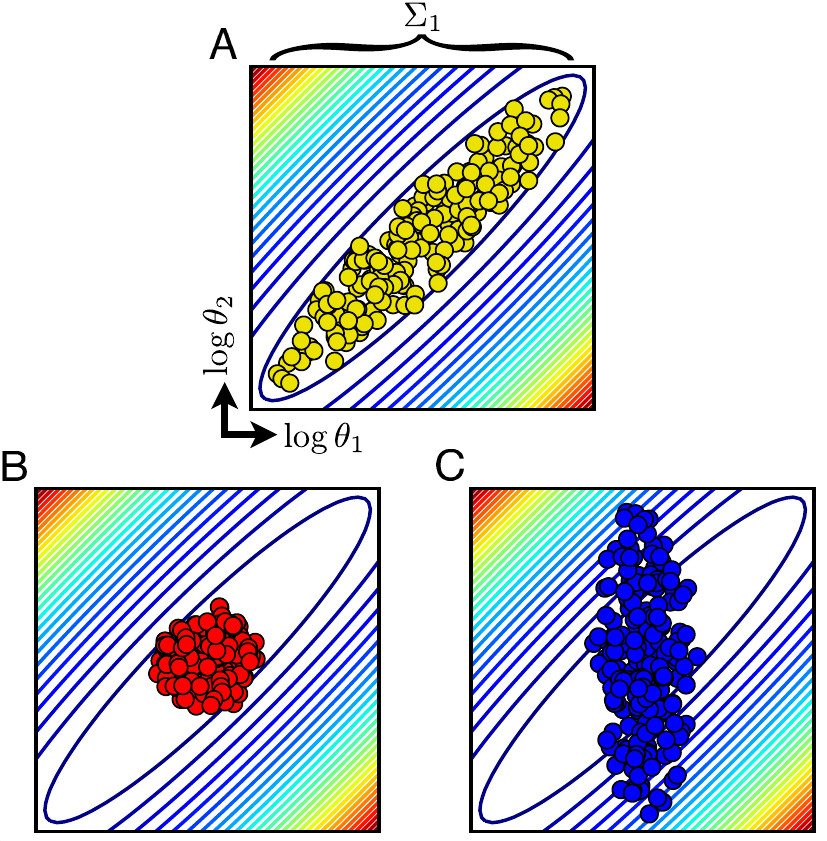} 
\caption{Sloppiness and uncertainties\\
As in Figure~\ref{fig:sloppiness}A, the contours
represent surfaces of constant model behavior deviation. The clouds of points
represent parameter set ensembles.\\
A: Collective fitting of model parameters
naturally constrains the parameter set ensemble along stiff directions and
allows it to expand along sloppy directions. The resulting ensemble may be very
large, yet encompass little variation in model behavior, yielding small
prediction uncertainties despite large parameter uncertainties.  ($\Sigma_1$
denotes the 95\% confidence for the value of $\theta_1$.)\\
B: If all parameters
are directly measured to the same precision, the parameter set ensemble is
spherical. The measurement precision required for well-constrained predictions
is set by the stiffest direction.\\
C: If one parameter (here $\theta_2$) is
known less precisely than the rest, the cloud is ellipsoidal.  If not aligned
with a sloppy direction, the cloud will admit many model behaviors and yield
large prediction uncertainties.  (Note that the aspect ratio of the real
contours can be greater than 1000.) }
\label{fig:explanation}
\end{center}
\end{figure}

The difficulty of extracting precise parameter values from collective fits in systems biology modeling is well-known~\cite{bib:Gadkar2005}.
%It is well known that collective fits of parameters in systems biology models often yield large parameter uncertainties~\cite{XXX}. 
Sloppiness offers an explanation for this
and predicts that it will
be true even for fitting to complete data that the model can fit perfectly.  In a collective fit, the parameter set
ensemble samples from all sets of parameters for which the model behavior is
consistent with the data.  Because sloppy models are very insensitive to
parameter combinations that lie along sloppy directions, the parameter set
ensemble can extend very far in those directions, as illustrated schematically
in Figure~\ref{fig:explanation}A.  As a result, individual parameters can be
very poorly determined (\emph{e.g.}, confidence interval indicated by $\Sigma_1$ in
Figure~\ref{fig:explanation}A).  Below we discuss a test of this prediction over all the models
in our collection. 

Unless one has direct interest in the kinetic constants for the underlying
reactions, uncertainties in model predictions are generally more
important than uncertainties in model parameters. The parameter set ensemble illustrated in
Figure~\ref{fig:explanation}A yields large uncertainties on individual
parameters, but can yield small uncertainties on predictions.  While the
fitting process allows the ensemble to expand along sloppy directions, the
fit naturally constrains the ensemble along stiff directions, so that model behavior varies
little within the ensemble, and predictions can be consequently tight.

Direct parameter measurements, on the other hand, will have uncertainties that
are uncorrelated with the model's underlying stiff and sloppy directions.  For
example, if all parameter measurements are of the same precision, the parameter
set ensemble is spherical, as illustrated in Figure~\ref{fig:explanation}B.
For tight predictions, this ensemble must not cross many contours, so the
required precision is set by the stiffest direction of the model.
Consequently, high precision parameter measurements are required to yield tight
predictions.  Moreover, these measurements must be complete.  If one parameter
is known less precisely, the parameter set ensemble expands along that
parameter axis, as illustrated in Figure~\ref{fig:explanation}C.  If that axis
is not aligned with a sloppy direction, model behavior will vary dramatically
across the parameter set ensemble and predictions will have correspondingly
large uncertainties.  Below we discuss tests of both these notions, exploring the effects of
direct parameter measurement uncertainty on predictions of a particular model.

\subsubsection{Parameter Values from Collective Fits}\label{sec:params_from_fits}

\begin{figure}
    \begin{center}
    \includegraphics{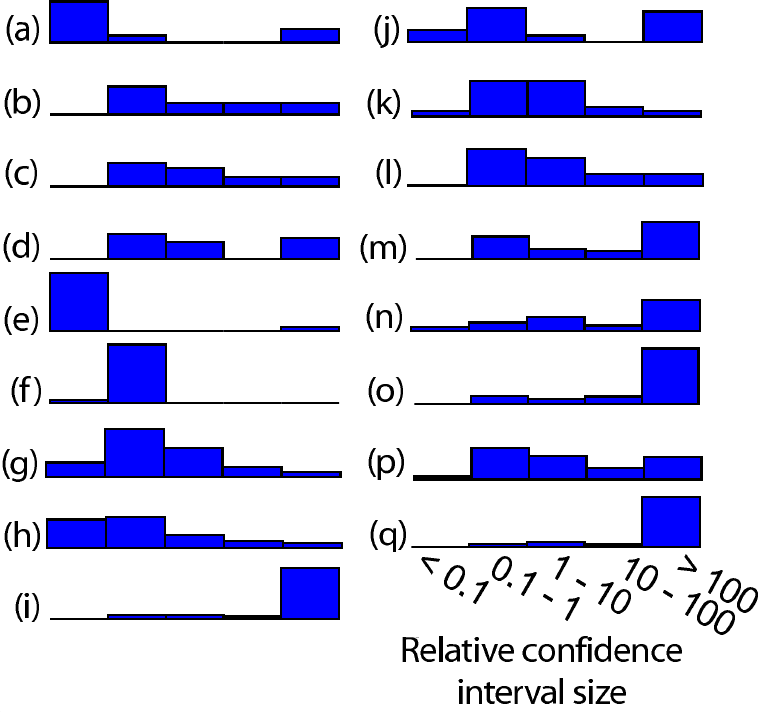} 
    \caption{Fitting parameters to idealized data\\
Shown are histograms of the relative confidence interval size $\Sigma$ for each
parameter in each model of our collection, after fitting 100 times as many
time-series data points (each with 10\% uncertainty) as parameters.  In most
cases a large number of parameters are left with greater than 100\%
uncertainty.  (A parameter constrained with 95\% probability to lie between 1
and 100 would have $\Sigma \approx 100$.)\\
Labels are as in Figure~\ref{fig:sloppiness}.}\label{fig:fitting_params} 
    \end{center}
\end{figure}

Does the sloppiness of these models really prevent one from extracting
parameters from collective fits?  Here we discuss a test of this prediction using an idealized fitting procedure.

Our \ChiSq measure of model behavior change (Equation~\ref{eqn:ChiSq})
corresponds to the cost function for fitting model parameters to 
continuous time-series data that the model fits perfectly at parameters
$\theta^*$; \HchiSq is the corresponding Fisher information matrix
(Equation~\ref{eqn:HchiSq}).  We used this idealized situation to test the
prediction that collective fits will often poorly constrain individual
parameters in our collection of sloppy models.

We defined the relative 95\% confidence interval size $\Sigma_i$ as the ratio
between parameter $i$ at the upper and lower extremes of the interval, minus
one.  (A parameter value constrained after the fit to lie between 10 and 1000
would have $\Sigma \approx 100$, while one constrained between 1.0 and 1.5 would have
$\Sigma = 0.5$.) We assumed 100 times as many data points (each with 10\% uncertainty) as the number of parameters in each model. Figure~\ref{fig:fitting_params} shows histograms of the
quadratic approximation to $\Sigma$ for each parameter in each model after
fitting such data. (See Methods.) For most of the models, the figure indicates that such fitting leaves many parameters with
greater than 100\% uncertainty ($\Sigma > 1$).
Indeed, even fitting this large amount of
ideal data can leave many parameter values very poorly determined, as
expected from the sloppiness of these models and our discussion of
Figure~\ref{fig:explanation}A.

The fact that nonlinear multiparameter models often allow a wide range of correlated parameters to fit data has long been appreciated. As one example, a 1987 paper by Brodersen \emph{et~al}.\ on ligand binding to hemoglobin and albumin empirically found many sets of parameters that acceptably fit experimental data, with individual parameter values spanning huge ranges~\cite{bib:Brodersen1987}.
Our sloppy model perspective (\cite{bib:Brown2003a,bib:Brown2004,bib:Waterfall2006}, Figure 1) shows that there is a deep underlying universal pattern in such least-squares fitting.
Indeed, an analysis of the acceptable binding parameter sets from the 1987 study shows the same characteristic sloppy eigenvalue spectrum we observed in Figure~\ref{fig:sloppiness}B (Supporting Text~S5).

\subsubsection{Predictions from Direct Parameter Measurements}\label{sec:pred_from_meas}

\begin{figure}
\begin{center} 
\includegraphics{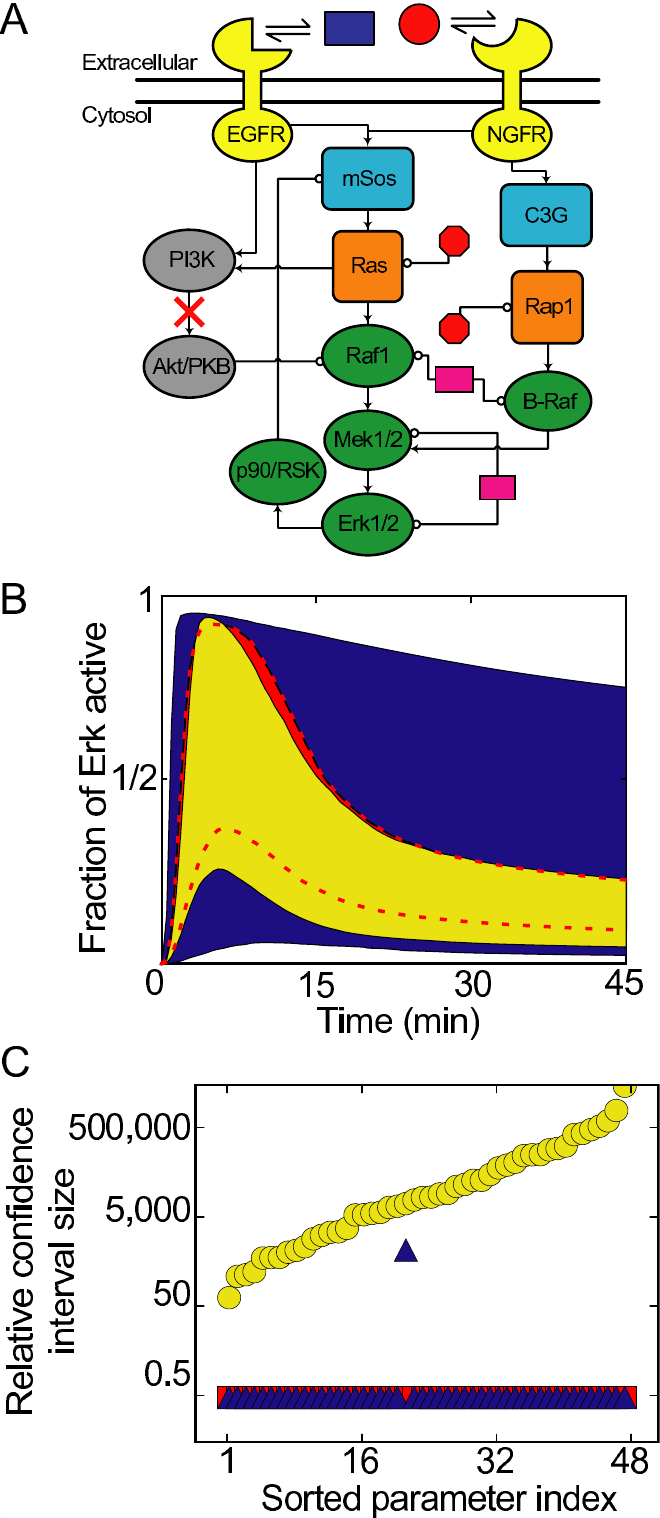} 
\caption{Parameter and prediction uncertainties\\
A: Our example prediction is for ERK activity upon EGF stimulation given PI3K
inhibition in this 48-parameter model of growth-factor-signaling in PC12
cells~\protect\cite{bib:Brown2004}.\\
B: Shaded regions are 95\% confidence intervals calculated via exhaustive Monte Carlo for our example 
prediction given various scenarios for constraining parameter values.\\
C: Plotted is the relative size $\Sigma$ of the 95\%
confidence interval for each parameter.\\
The scenarios represented are: (red,
squares) all model parameters individually measured to high precision, (blue,
triangles) all parameters precisely measured, except one estimated to low
precision, (yellow, circles) all parameters collectively fit to 14 real cell-biology
experiments.  Precisely measured individual parameter values enable a tight
prediction (B: middle red band), but even one poorly known parameter can
destroy predictive power (B: wide blue band).  In contrast, the collective fit
yields a tight prediction (B: tightest yellow band) but only very loose
parameter constraints (C: circles).
The large parameter uncertainties from the collective fit (C: circles) calculated here by Monte Carlo are qualitatively similar to those seen in the linearized fit to idealized data (Figure~\ref{fig:fitting_params}, model (i)). (For clarity, the dashed red lines trace the boundary of the red confidence interval.)
}\label{fig:uncerts} 
\end{center} 
\end{figure}

Figures~\ref{fig:explanation}B and~\ref{fig:explanation}C suggests that direct parameter measurements must be both precise and complete to usefully constrain predictions in sloppy systems. Here we discuss a test of this notion in a specific model.

We worked with the
48-parameter growth-factor-signaling model of \BrownEtAl, shown schematically
in Figure~\ref{fig:uncerts}A~\cite{bib:Brown2004}.  The parameters in this
model were originally collectively fit to 14 time-series cell-biology
experiments.  We focused on this model because it is instructive to compare
our results concerning direct parameter measurements with prior results from
collective fitting.  For our analysis, we assumed that hypothetical direct
parameter measurements would be centered about the original best-fit values.

One important test of the model was a prediction of the time-course of ERK
activity upon EGF stimulation, given inhibition of the PI3K branch of the
pathway.  The yellow shaded region in Figure~\ref{fig:uncerts}B shows the
uncertainty bound on this prediction from the original collective
fit, calculated by exhaustive Monte Carlo~\cite{bib:Brown2004}.  The tightness of this prediction is remarkable
considering the huge uncertainties the collective fit left in the individual
parameter values (yellow circles in Figure~\ref{fig:uncerts}C).  Not a single
parameter was constrained to better than a factor of 50.

How precise would direct parameter measurements have to be to yield as tight a
prediction as the collective fit?  For this prediction, the PI3K branch
(inhibited) and C3G branch (NGF-dependent) of the pathway are irrelevant in the
model; the remaining reactions involve 24 parameters.  To achieve the red
prediction in Figure~\ref{fig:uncerts}B, all 24 involved parameters must be
measured to within a factor of plus or minus 25\% (Figure~\ref{fig:uncerts}C,
red squares).  With current techniques, measuring even a single \emph{in vivo}
biochemical parameter to such precision would be a challenging experiment.
Such high precision is required because, as illustrated in
Figure~\ref{fig:explanation}B, the measurements need to constrain the stiffest
combination of model parameters.

What if a single parameter is left unmeasured?  For example, consider high
precision measurements of 23 of the 24 involved parameters, all but the rate
constant for the activation of Mek by Raf1.  For this unmeasured parameter, we
assumed that an informed estimate could bound it at 95\% confidence to  within
a total range of 1000 (\emph{e.g.}, between $1 s^{-1}$ and $1000 s^{-1}$).  The resulting prediction
(blue in Figure~\ref{fig:uncerts}B) has very large uncertainty and would likely
be useless.  Note that these hypothetical measurements constrain every
individual parameter value more tightly than the original collective fit (blue
triangles versus yellow circles in Figure~\ref{fig:uncerts}C), yet the
prediction is much less well-constrained.  Neither this parameter nor this
prediction is unique.  Uncertainty for this prediction is large if any one of
about 18 of the 24 involved parameters is unmeasured (Supporting Text S2). Furthermore,
other possible predictions in this model are similarly fragile to single
unmeasured parameters (Supporting Text S3).

To usefully constrain Brown \emph{et~al}.'s model, direct parameter measurements would need to be both precise and complete. By contrast, collective parameter fitting yielded tight predictions with only a modest number of experiments. These results are expected given the model's sloppiness.

\section{Discussion}
By examining seventeen models from the systems biology literature~\citecoll,
we showed that their parameter sensitivities all share striking common
features deemed `sloppiness'; the sensitivity eigenvalues span many 
decades roughly evenly (Figure~\ref{fig:sloppiness}B), and tend not to
be aligned with single parameters (Figure~\ref{fig:sloppiness}C).
We argued that sloppy parameter sensitivities help explain the difficulty of extracting precise parameter estimates from collective fits, even from comprehensive data. 
Additionally, we argued that direct parameter measurements should be inefficient at constraining predictions from sloppy models.
We then showed that collective parameter fits to complete time-series
data do indeed yield large parameter uncertainties in our model collection 
(Figure~\ref{fig:fitting_params}). 
Finally, we confirmed for the
48-parameter signaling model of \BrownEtAl~\cite{bib:Brown2004} that 
direct parameter measurements must be formidably precise and complete
to usefully constrain model predictions (Figure~\ref{fig:uncerts}).

What causes sloppiness? 
(1)~Fundamentally, sloppiness involves an extraordinarily singular coordinate transformation in parameter space between the bare parameters natural in biology (\emph{e.g.}, binding affinities and rate constants) and the eigenparameters controlling system behavior, as discussed in~\cite{bib:Waterfall2006}.
Both experimental interventions and biological evolution work in the bare parameter space, so this parameterization is fundamental to the system, not an artifact of the modeling process.
(2)~Sloppiness depends not just upon the model, but also on the data
it is fit to; exhaustive experiments designed to decouple the system and
separately measure each parameter will naturally not yield sloppy
parameter sensitivities. 
(3)~In biological systems fit to time-series data, Brown and Sethna~\cite{bib:Brown2003a} note that sloppiness may arise due to 
under-determined systems, proximity to bifurcations, and separation of time or concentration scales, but they doubt that these can explain all the sloppiness found in their model.
Our analysis includes complete data on all species, and hence is overdetermined. 
Small eigenvalues near bifurcations are associated with 
dynamic variables, and not the system parameters we investigate. 
To study the effect of time and concentration scales we calculated \HchiSq for a version of the \BrownEtAl model in which all concentrations and rate constants were scaled to one~\cite{bib:SiggiaNote}. The resulting model remains sloppy, with eigenvalues roughly uniformly spanning five decades (Supporting Text S4).
%We are not certain of the underlying cause of the remaining sloppiness.
(4)~Motivated by simple example systems, we have argued elsewhere that sloppiness emerges from a redundancy between the effects of different parameter combinations~\cite{bib:Waterfall2006}.
We are presently investigating decompositions of parameter space
into sloppy subsystems~\cite{bib:WaterfallPhD} and the use of physically or biologically motivated nonlinear coordinate changes
to remove sloppiness or motivate simpler models.
These potential methods for model refinement, however, demand a
complete and sophisticated understanding of the system that is
unavailable for many biological systems of current interest.

%Sloppiness may also provide a more quantitative understanding of robustness~\cite{bib:Wagner2005} in certain contexts;
%we too find that our system behavior is remarkably independent of parameter
%choice, depending on only three or four stiff directions rather than 
%forty or fifty independent parameter constraints.

Parameter estimation has been a serious obstacle in systems biology modeling.
With tens of unknown parameters, a typical modeling effort might draw some values
from the literature (possibly from \emph{in vitro} measurements or different cell lines)~\cite{bib:Kholodenko2000, bib:Curto1998},
set classes of constants to the same value (\emph{e.g.}, phosphorylation rates)~\cite{bib:Vilar2002, bib:Edelstein1996, bib:Sasagawa2005}, 
and adjust key parameters to qualitatively best fit the existing data~\cite{bib:Ueda2001,bib:Chen2004, bib:Locke2005}.
In retrospect, these approaches may be successful because the
models are sloppy---they can be tuned to reality by adjusting one key parameter per stiff direction, independently of how reliably the other parameters are estimated. 

Computational modeling is a potentially invaluable tool for extrapolating from current experiments and distinguishing between models.
But we cannot trust the predictions of these models without testing how much they depend on uncertainties in these estimated parameters.
Conversely, if we insist upon a careful uncertainty analysis, it would seem unnecessary to insist upon tight prior estimates of the parameters, since they do not significantly enhance model predictivity.
Because the behavior of a sloppy model is dominated by a few stiff directions that nonetheless involve almost all the parameters, direct parameter measurements constrain predictions much less efficiently than comparably difficult experiments probing collective system behavior.

Our suggestion of making predictions from models with very poorly known parameters may appear dangerous.
A model with tens or hundreds of unmeasured parameters might seem completely untrustworthy; we certainly believe that any prediction derived solely from a best-fit set of parameters is of little value.
Uncertainty bounds derived from rigorous sensitivity analysis, however, distinguish those predictions that can be trusted from those that cannot.
Of course, successful fits and predictions may arise from models that are incorrect in significant ways; for example, one model pathway with adjusted parameters may account for two parallel pathways in the real system.
A model that is wrong in some details may nevertheless be useful in guiding and interpreting experiments.
For computational modeling to be useful in incompletely understood systems, we must focus not on building the final, perfect, model with all parameters precisely determined, but on building incomplete, tentative and falsifiable models in the most expressive and predictive fashion feasible.

Given that direct parameters measurements do not efficiently constrain model
behavior, how do we suggest that experimentalists decide what experiment
to do next? If the goal is to test the assumptions underlying a model, 
one should look for predictions with tight uncertainty estimates that can
be readily tested experimentally. If the goal is to reduce uncertainty in
crucial model predictions, one must invoke the statistical methods of
optimal experimental design, which we have studied 
elsewhere~\cite{bib:Casey2006} and which may be conveniently implemented
in modeling environments that incorporate sensitivity analysis (such as
SloppyCell~\cite{bib:SloppyCell}). 

In our approach, the model and its parameters cannot be treated in isolation from the data that informed model development and parameter fitting.
This complicates the task of exchanging knowledge in the modeling community.
To support our approach, standards such as SBML~\cite{bib:Hucka2003} that facilitate automated model
exchange will need to be extended to facilitate automated data exchange.

Every one of the 17 systems biology models we studied exhibits a sloppy
spectrum of parameter sensitivity eigenvalues; they all span many decades roughly
evenly and tend not be aligned with single parameters.  This
striking and apparently universal feature has important consequences for the modeling process.  It suggests that modelers would be wise to try collective parameter fits and to focus not on the quality of their parameter values but on the quality of their predictions.

\section{Methods}\label{sec:methods} 
\subsection{Hessian Computations} 
\HchiSq can
be calculated as
\begin{equation} \HchiSq_{j,k} = \frac{1}{N_c\,N_s} \sum_{s,
c}\frac{1}{T_c\,\sigma_s^2} \int^{T_c}_0 {\frac{d \, y_{s,c}(\theta^*,
t)}{d \log \theta_j}} {\frac{d\, y_{s, c}(\theta^*, t)}{d \log \theta_k}} \bigg|_{\theta^*}\,\mathrm{d}t.\label{eqn:HchiSqEx}
\end{equation}
Second derivative terms $\left({d^2\, y_{s,c}(\theta^*,
t)}/{d \log \theta_i \, d \log \theta_j}\right)$ might be
expected, but they vanish because we evaluate \HchiSq at $\theta^*$.
Equation~\ref{eqn:HchiSqEx} is convenient because the first derivatives
${d\, y_{s, c}(\theta^*, t)}/{d \log \theta_k}$ can be calculated by
integrating sensitivity equations.  This avoids the use of finite-difference
derivatives, which are troublesome in sloppy systems.

The projections $P_i$ of the ellipsoids shown in
Figure~\ref{fig:explanation}A onto bare parameter axis $i$ are proportional to
$\sqrt{\big(\operatorname{inv}\HchiSq\big)_{i,i}}$.  The intersections
$I_i$ with axis $i$ are proportional to $\sqrt{1/\HchiSq_{i,i}}$,
with the same proportionality constant.

\subsection{Parameter Uncertainties}\label{sec:methods_fit}
To rescale \HchiSq so that it corresponds
to fitting $N_d$ data points, each with uncertainty a fraction $f$ of the
species' maximal value, we multiply \HchiSq by $N_d/f^2$.  In the quadratic
approximation, the one-standard-deviation uncertainty in the logarithm of
parameter $\theta_i$ after such a collective fit is given by $\sigma^2_{\log
\theta_i} = \left(f^2/N_d\right) \big(\operatorname{inv}\HchiSq\big)_{i,i}$.  The relative
size of the 95\% confidence interval is then $\Sigma_{i} = \exp\left(4\sigma_{\log
\theta_i}\right) - 1$.

\subsection{Prediction Uncertainties}\label{sec:methods_pred_uncerts}
The red and blue prediction uncertainties
shown in Figure~\ref{fig:uncerts}B were calculated by randomly generating 1000
parameter sets consistent with the stated parameter uncertainties.  (For each
parameter $\theta_i$, the logarithm of its value is drawn from a normal
distribution with mean $\log \theta_i$ and standard deviation $\sigma_{\log
\theta_i}$ specified by desired $\Sigma$.) For each parameter set, the Erk time
course was calculated, and at each timepoint the shaded regions in the figure
contain the central 95\% of the time courses.

\subsection{Software} 
All computations were performed in the open-source
modeling environment SloppyCell, version 0.81~\cite{bib:SloppyCell}. SBML files
and SloppyCell scripts to reproduce all presented calculations are in Dataset S1.

\section{Supporting Information} 
Text S1: Stiffest Eigenvectors

Text S2: Effect of Other Poorly Determined Parameters

Text S3: Fragility of Other Predictions

Text S4: Rescaled Model of \BrownEtAl

Text S5: Eigenvalue Analysis of Brodersen \emph{et al.}\ Binding Studies

Dataset S1: SBML Files, SloppyCell Scripts, and \ChiSq-Hessians

\subsection{Accession Numbers}
Models discussed that are in the BioModels database~\cite{bib:BioModels} are: 
(a)~BIOMD0000000005,
(c)~BIOMD0000000003,
(d)~BIOMD0000000035,
(e)~BIOMD0000000002,
(f)~BIOMD0000000010,
(h)~BIOMD0000000021,
(i)~BIOMD0000000033,
(k)~BIOMD0000000022,
(l)~BIOMD0000000055,
(n)~BIOMD0000000015,
(o)~BIOMD0000000051,
(p)~BIOMD0000000056,
(q)~BIOMD0000000049.

\section{Acknowledgments}
We thank Eric Siggia for suggesting study of the rescaled model of \BrownEtAl 
We also thank Rick Cerione and Jon Erickson for sharing their biological insights and John Guckenheimer, Eric Siggia, and Kelvin Lee for helpful discussions about dynamical systems.
Computing resources were kindly provided by the USDA-ARS plant pathogen systems biology group in Ithaca, NY.
Finally, we thank several anonymous reviewers whose comments strengthened the manuscript.

\section{Funding}
RNG was supported by an NIH Molecular Biophysics Training Grant, T32-GM-08267.
JJW was supported by a DOE Computational Science Graduate Fellowship.
CRM acknowledges support from USDA-ARS project 1907-21000-017-05.
This work was supported by NSF grant DMR-0218475.

\bibliography{sloppy}

%\clearpage
%\listoffigures

\end{document}